# Recovery of entire shocked samples in a range of pressure from ~100 GPa to Hugoniot Elastic Limit

Keita NAGAKI[1], Toshihiko KADONO[2*], Tatsuhiro SAKAIYA[1], Tadashi KONDO[1], Kosuke KUROSAWA[3], Yoichiro HIRONAKA[4], Keisuke SHIGEMORI[4], and Masahiko ARAKAWA[5]

[1] *Graduate School of Science, Osaka University, 1-1 Machikaneyama, Toyonaka, Osaka 560-0043, Japan*

[2] *School of Medicine, University of Occupational and Environmental Health, 1-1 Iseigaoka, Yahata, Kitakyusyu 807-8555, Japan*

[3] *Planetary Exploration Research Center, Chiba Institute of Technology, 2-17-1 Tsudanuma, Narashino, Chiba 275-0016, Japan*

[4] *Institute of Laser Engineering, Osaka University, 2-6 Yamadaoka, Suita, Osaka 565-0871, Japan*

[5] *Graduate School of Science, Kobe University, 1-1 Rokkodai, Nada, Kobe 657-8501, Japan*

\* Corresponding author: T. Kadono

*E-mail address:* kadono@med.uoeh-u.ac.jp

**Abstract** - We carried out laser shock experiments and wholly recovered shocked olivine and quartz samples. We investigated the petrographic features based on optical micrographs of sliced samples and found that each recovered sample comprises three regions, I (optically dark), II (opaque) and III (transparent). Scanning electron microscopy combined with electron back-scattered diffraction shows that there are no crystal features in the region I; the materials in the region I have once melted. Moreover, numerical calculations performed with the iSALE shock physics code suggest that the boundary between regions II and III corresponds to Hugoniot Elastic Limit (HEL). Thus, we succeeded in the recovery of the entire shocked samples experienced over a wide range of pressures from HEL (~ 10 GPa) to melting pressure (~ 100 GPa) in a hierarchical order.

# INTRODUCTION

Hypervelocity impacts generate high pressures and shock waves in target materials. After the passage of the shock waves, the physical and/or chemical properties of the target materials are often substantially altered due to energy deposition; phase transitions or transformations such as vaporization and melting (e.g., Ahrens and O'Keefe 1972) and shock-metamorphic effects in rocks and minerals (e.g., Stöffler 1972) occur. The amount of altered target-materials has been paid attention, for example the extent of vaporization and melting (e.g., Kieffer and Simmonds 1980; Pierazzo et al. 1997), the volume of the materials experienced with higher pressures than Hugoniot Elastic Limit (HEL, the limiting value of stress in solid media beyond which plastic or irreversible distortions occur), which would affect the morphology of final craters such as central pits, terraces, and rings (e.g., Ferriére et al. 2008), and the largest fragment mass in impact fragmentation (e.g., Mizutani et al. 1990; Mitani 2003).

The recovery and analysis of shocked samples in laboratories have been one of the important methods to study the effect of shock wave passing in materials. In previous sample recovery experiments, the shock-metamorphic effects such as phase transformations and shock-induced petrographic features have been investigated mainly from thin samples shocked at constant pressures in metal containers (see a review Stöffler and Langenhorst 1994) and sometimes from the highly shocked materials ejected by an excavation flow in cratering experiments, which are collected using ejecta catchers (e.g., Stöffler et al. 1975; Ebert et al. 2013). However, spatial distributions of shocked materials in rather large pressure ranges (e.g., those sufficient to melt down below the HEL; for olivine, pressure range from > ~ 80 GPa below ~ 10 GPa) have not been reported previously in shock experiments involving minerals; such distributions should be necessary to quantitatively know the amount of shock-metamorphosed materials and to observe the transitions between various shock stages, which would be used to calibrate some shock effects to pressures after the passage of a decaying shock wave.

In this paper, we experimentally recover entire samples shocked in a large pressure range from higher than melt down below HEL and discuss the relation between the petrographic features of the recovered samples and the pressures experienced. As mineral samples, we used olivine and quartz; they are planetary major components, and abundant in the data obtained by the recovery and analysis of shocked samples (for olivine, e.g., Reimold and Stöffler 1978; Syono et al. 1981a; Ohtani et al. 2004; Tschauner et al. 2009; and for quartz, e.g., Stöffler and Langenhorst (1994); a recent series of experiments using sandstones (e.g., Kowitz et al. 2013a)).

To obtain high shock pressures, we used a laser facility. Recently, hypervelocity impacts have been experimentally simulated at pressures exceeding a few hundred GPa, corresponding to impact velocities > 10 km/s using high-power lasers (e.g., Kadono et al. 2010; 2012; Kurosawa et al. 2010; 2012a, b; Takasawa et al. 2011; Ohno et al. 2014). Here, we also generated pressures > a few hundred GPa by using a high-power laser, the GEKKO-XII HIPER laser at the Institute of Laser Engineering in Osaka University.

Numerical modeling has been carried out to provide the explanation for the shock processes in impact experiments, including sample recovery experiments (e.g., Kowitz et al. 2013b). In this paper, we also carried out a numerical calculation simulating hypervelocity impacts to discuss the relation between the petrographic features of the recovered samples and the pressures experienced. Based on this discussion, our goal is to find some petrographic features of recovered samples as an indicator of melting and

HEL.

## EXPERIMENTAL AND ANALYTICAL METHODS

Figure 1 is a schematic of the recovery experiments. The samples were single-crystal San Carlos olivine (Fo89) and artificially manufactured single crystal quartz. The cubic samples of 3 x 3 x 3 mm dimension were surrounded by aluminum (Al) cases except for the laser irradiation surface, which was covered with a titanium (Ti) sheet of 0.2-0.6 mm thickness to prevent the ejection of shocked samples from the Al cases. Reflection from the surrounding Al cases should exert smaller effects than that from stainless steel cases, which have been frequently adopted in previous recovery experiments, because the shock-impedance of the samples is similar to that of Al (e.g., Melosh 1989) in comparison with stainless steel.

There are two types in laser experiments simulating hypervelocity impacts. One is flyer impact method. Laser irradiates spherical or sheet flyers directly or ablator (plastics or metals) attached to the flyers and generates plasma of high pressure and temperature, which rapidly expands to accelerate the flyers as rockets. Under appropriate laser conditions, the flyers in condensed phase are accelerated to a velocity higher than ~ 10 km/s and impact on targets (e.g., Kadono et al. 2010; 2012). The second one is direct drive method; there is no flyer. The laser directly irradiates ablators attached to target samples, and ablation plasma of extremely high pressure produces a strong shock in the samples (e.g., Kurosawa et al. 2012a). In this method, the achievable peak shock pressure is much higher, and the experimental setup is rather simpler than the flyer method.

In this paper, we adopted the second (direct drive) method for the experimental setup to be simpler. The Ti sheet was irradiated with a laser (energy and wavelength of 0.7 kJ – 4.4 kJ and 1053 nm, respectively). The temporal pulse shape was approximately triangle; the foot-to-foot pulse duration was ~24 ns, and the peak was around 6 ns from the beginning of the pulse (Fig. A1a in the Electronic Appendix). The spatial distribution was approximately Gaussian and the full width at half maximum spot size was about 500 μm for olivine or 400 μm for quartz. Throughout the experiments, 4 shots for olivine and 2 shots for quartz were carried out. Based on a simple relationship between ablation plasma pressure $P$ and laser intensity $I$: $P \sim I^{2/3}$ (e.g., Fabbro et al. 1985; Atzeni and Meyer-Ter-Vehn 2004), we define an average pressure $P_0$ of ablation Ti plasma as $P_0 = K(E/S)^{2/3}$, where $E$ is total laser energy, $S$ is laser-irradiation spot area, and $K$ is a calibrated proportional-coefficient for the above laser irradiation conditions. The calibration experiment reveals that the pressure of the ablation plasma is approximately regarded as a constant during the laser irradiation (Fig. A2 in Electronic Appendix). Based on this constant pressure, we evaluated $K$ in the above equation. In Table I, the experimental condition, sample, the thickness of Ti, total laser energy $E$, laser irradiation spot diameter, and estimated $P_0$, are shown.

After the laser irradiation, we recover the entire shocked samples. Most parts of the Ti sheets were ejected and not recovered. Slices of the samples attached to glass plates were obtained as follows. The recovered samples were solidified with epoxy-resin and sliced with a diamond blade. The surface of one sample fragment was polished and attached to a glass plate, and again sliced with the diamond blade. Finally, the sample surface was polished to a sample thickness of ~0.3 mm.

We analyzed these slices of the recovered samples using optical microscopes to observe the petrographic features. Then, the slices were coated with carbon and observed with a field emission-scanning electron microscope (FE-SEM; JSM-7001F,

JEOL) at Osaka University and analyzed with Electron Back Scatter Diffraction Patterns (EBSD; Nordlys, HKL) system. The working distance was 9.1 mm and the acceleration voltage was 15 kV.

## RESULTS OF ANALYSIS OF RECOVERED SAMPLES

Figure 2a shows a slice of the recovered olivine sample irradiated with $P_0$ ~190 GPa laser ablation pressure (shot no. 33757). The shock wave proceeded from the right-hand side of the section. The right-hand edge of the sample that contacted with the Ti sheet was concave about 85 μm from the original surface (most parts of the Ti sheet were ejected and lost). Figure 2b shows the intensity (the degree of transparency) distribution in the long white rectangle as a function of the distance from the Ti surface. Based on the intensity levels, we can distinguish three regions, (Region I) the lowest intensity region near the surface, (Region II) the middle region, and (Region III) the brightest (transparent) region far from the surface. Each boundary between these regions, where the intensity changes largely, is defined as the middle point between the edge and base of the plateau in the intensity, indicated by adjacent, thin vertical lines in Fig. 2b (the plateaus correspond to these regions). The positions of these boundaries for each shot are shown in Table 1 as the distance from the Ti surface.

Figure 2c is a SEM image of the area delineated by the white rectangle in Fig. 2a, farthest to the surface of the sample (Region III) in Fig. 2a, which is characterized by fewer cracks in closer alignment, while the SEM image of Region II delineated by the second white rectangle in the Fig. 2a is etched with randomly oriented cracked (Fig. 2d). Figure 2e is the closest rectangle to the surface of the sample in Fig. 2a. The right-hand side of the image (the shallower parts of the sample) is characterized by a low crack density and relatively flat areas (Region I). In addition, EBSD analyses show no clear crystal structure of the materials in Region I (Fig. 2f: taken from the shallower parts of Fig. 2e). On the other hand, the left-hand side (the deeper parts) in Fig. 2e (Region II) is heavily cracked, and EBSD image (Fig. 2g: taken from the deeper parts of Fig. 2e) is unlike those in Fig. 2f (Region I). These results suggest that the materials in Region I have been melted and rapidly solidified. Thus, Region I, which is nearest to the Ti sheet, is expected to be a melt region, though more detailed analyses of Region I should be carried out to find definite evidences of melting such as vesicles or flow features.

In Fig. 3, optical (transmitted light) microscope slice images of all shots are shown.

For olivine samples (Figs. 3a-3d), the similar features in the intensity distribution as seen in Fig. 2a can be recognized in every image, and, hence, we can distinguish the regions corresponding to Region I – III in Fig. 2a (there seems no Region I in Fig. 3a the lowest $P_0$ case). As the initial pressures increase, the spatial ranges of Region I and II increase. For quartz samples (Figs. 3e and 3f), transmitted and reflected light images are shown. Though Region I is recognized, Region II is not clear in the transmitted light images while, in the reflected light images, Region II is clearly recognized. Thus, in both materials, the slices of recovered samples can be divided into three regions based on the optical microscope images.

## NUMERICAL SIMULATION

To evaluate the pressure experienced in the recovered samples, we carried out numerical simulations using the shock physics code iSALE-2D (Wünnemann et al. 2006). iSALE-2D is an extension of the SALE code (Amsden et al. 1980), which is capable of modeling shock processes in geologic materials (Melosh et al. 1992; Ivanov et al. 1997; Collins et al. 2004).

We consider a geometrical case in axial symmetry. Target was cylindrical dunite or quartz (3 mm in height and 1.5 mm in radius) attached with a Ti sheet on the top flat surface of the cylindrical targets. The thickness of the Ti sheet was changed according to the experimental conditions (0.2-0.6 mm in thickness). The target surfaces except for the one covered with the Ti sheet were surrounded by Al cases as the experiments with a thickness of 0.25 mm.

The calibration experiment showed that the laser ablation pressure was approximately regarded as a constant (Fig. A2 in Electronic Appendix). Hence, we assume a constant initial shock-pressure (a square pulse) in the iSALE simulations. To represent a square pulse, though there was no flyer in our laser direct irradiation experiments, we set "impact" conditions as follows. A cylindrical Ti projectile was virtually considered and assumed to impact a Ti sheet covering the sample. Impact velocity was set to generate the average ablation pressures $P_0$ generated in the Ti sheet, which were ~90-430 GPa in our experiments (Table 1), corresponding to an impact velocity of ~5-15 km/s. The dimensions of the virtual projectiles are set based on the laser conditions. The radius of the projectile was the same as the laser spot radius, 0.25 mm for dunite and 0.2 mm for quartz, respectively. The thickness (height) was determined such that the laser irradiation time of 24 ns is equal to the sum of the travel times of the shock wave caused by the impact to the rear surface of the projectile and the rarefaction wave generated at the rear surface of the projectile back to the boundary between the projectile and target (it is noted that pressure attenuation profiles change as a function of the height of the projectile, but the dependence is gradual as shown in Fig. A3 of Electronic Appendix). According to our initial average ablation pressures, the height was changed 0.15-0.25 mm. The flat surface of the projectile normally impacted to the Ti sheet on the targets. Pressure profile generated by the impact at the point between the virtual projectile and the Ti sheet is shown in Fig. A1b of the Electronic Appendix.

The general setup for the calculations is described in Table A1 and the material parameters set in the calculations are listed in Tables A2 and A3 in Electronic Appendix. We used the Tillotson EOS (Tillotson 1962) for Ti and Al and ANEOS (Thompson and Lauson 1972) for dunite and quartz. The strength model (Ivanov et al. 1997) for dunite and quartz and the Johnson-Cook strength model (Johnson and Cook 1983) for Ti and Al were used, respectively. Note that the parameters for a popular titanium alloy

Ti$_6$Al$_4$V were used instead of pure Ti in the strength model (See Table A3 and Fig. A4 of the Electronic Appendix). For comparison, the simulations without strength were also carried out. We inserted Lagrangian tracer particles into each computational cell to investigate the peak pressure experienced as a function of their original positions. The end time of calculation was set to 1 μs, which is typically twice than the travel time of a shock wave in the targets along the central axis and enough to obtain the maximum experienced pressure.

To validate the code, we compare the results by iSALE with and without strength to previous models for dunite - dunite impact at 10 km/s obtained by Pierazzo et al. (1997) and Mitani (2003). The result by iSALE without strength is almost the same profile as the one by Pierazzo et al. (1997), and the result by iSALE with strength shows that the pressure decreases more quickly than that by Pierazzo et al. (1997) and is consistent with that by Mitani (2003), in which a strength model is included (Fig. A5 of the Electronic Appendix).

## DISCUSSION

In Fig. 4, the shock pressures obtained by iSALE with strength for the initial pressure of $P_0$ of 150 and 190 GPa are shown as a function of the distance normalized by the projectile radius. For comparison, we indicate the positions of the boundaries between Regions I and II and between II and III along the central axis, obtained from two recovered samples with $P_0$ of 150 and 190 GPa, as thin (I-II) and bold (II-III) vertical lines. The positions are normalized by the laser spot radius (0.25 mm for olivine). We also show the pressure of the onset of melting and HEL as horizontal lines. Various shock pressures of the onset of melting have been proposed (e.g., > 70 GPa for olivine and 50-65 GPa for quartz referred in Melosh (1989)). Here, we set the onset pressure to be 82 GPa for olivine and 70 GPa for quartz, both which are based on the concept of entropy matching between shocked and released states (Ahrens and O'Keefe 1972; Sugita et al. 2012). Note that this pressure is not the melting pressure in compressed, high-temperature states, which is higher, 130 GPa for olivine (Holland and Ahrens 1997) and 120 GPa for quartz (Akins and Ahrens 2002). The HEL for olivine is different depending on the crystal axis: ~12 GPa (010), 8.7 GPa (001) and ~6 GPa (100) (Syono et al. 1981b). Since the crystal axes of the samples are arbitrarily oriented in the experiments, we average the pressures at the HEL among the tri-axes, yielding

Voigt-Reuss-Hill average of 9 GPa. The HEL for quartz is set to be 9.5 GPa (Melosh 1989; Akins and Ahrens 2002).

The region I is suggested as that of the melting and should move from the original positions due to shock-driven particle movement (i.e., crater formation process). Noting that the results by iSALE in Fig. 4 show the maximum experienced pressures at the original positions, it is qualitatively consistent that the position of the boundary between the regions of I and II is slightly distant than the melting region obtained in the numerical results. On the other hand, the normalized positions $d/r_p$ of the boundary between II and III is very close to the HEL obtained in iSALE, suggesting that the boundary between II and III corresponds to HEL (the movement in the crater formation process at the depth around HEL would be smaller than the melt region).

Figure 5a shows a result by iSALE with strength as a pressure contour map. The corresponding experimental condition is that of the recovered sample shown in Fig. 2a. The two red curves indicate the onset of melting and HEL. Figure 5b shows the pressure contour in Fig. 5a superimposed on the image shown in Fig. 2a. It appears that the region surrounded by HEL agrees well with the region II in the recovered sample not only along the central axis as in Fig. 4 (we superimpose each slice image of the recovered olivine or quartz samples on the pressure contour map obtained by iSALE for corresponding experimental conditions in Fig. A6).

The pressure at the point between the Ti cover sheet and the sample surface on the central axis, $P_s$, obtained by the numerical simulations, is listed in Table 1. In all shots except for the shot no. 35245, $P_s$ is larger than the pressure of the onset of melting. Actually, we can recognize the region I in the slices of the recovered samples except for the shot no. 35245.

Thus, the comparison with the numerical results suggests that the boundary between II and III is corresponding to the HEL. Moreover, the SEM image of Region II is etched with randomly oriented cracked (Fig. 2d) while Region III (Fig. 2c: SEM image of the farthest rectangle in Fig. 2a) is characterized by fewer cracks in closer alignment. Consequently, the material loses its strength in Region II while it should retain its strength in some directions in Region III. Therefore, it can be said that the boundary between II and III corresponds to HEL and that the high number density of fine cracks generated in Region II is likely responsible for reduced transparency in that region as shown Figs. 2a and 2b.

Finally, we should note that the comparison between the recovered sample in Fig. 2a and the two-dimensional contour map in Fig. 5 suggests that the extent of the melting region (the thickness of the region I) seems almost the same, though the position of Region I is different. This implies that we can recover the most volume of impact-induced melt.

## CONCLUSION

We performed laser shock experiments and recovered shocked samples. We investigated the petrographic features of the recovered samples by means of optical microscopy and found that each sample is divided into three regions: I (optically dark), II (opaque), and III (transparent). The analyses of the samples using SEM and EBSD show that the boundary between I and II is corresponding to the onset pressure of melting. Moreover, the pressure attenuation profiles as a function of the distance obtained by numerical calculations suggest that the boundary between II and III is HEL. Thus, we succeeded in the recovery of the entire samples shocked to pressures from HEL (~ 10 GPa) to melt (~ 100 GPa) in the hierarchical order. The recovery of entire shocked samples and the conversion into experienced pressures from petrographic features would lead us to investigate the spatial distribution of the shock metamorphism successively as a function of pressure and provide information on the amount of altered materials in the future. Moreover, as an indicator of the experienced pressures such as HEL and melt, the intensity (the degree of transparency) distributions in the optical microscope images may be useful in the analyses of recovered natural samples.

*Acknowledgments* - This work was performed as Joint Research in the Institute of Laser Engineering, Osaka University. We are grateful to the GXII technical crew for their exceptional support during the experiments and A. Tsuchiyama, Y. Imai, T. Matsumoto, and J. Matsuno for their cooperation in the measurements by SEM. We appreciate the developers of iSALE, including G. S. Collins, K. Wünnemann, B. A. Ivanov, H. J. Melosh, and D. Elbeshausen. We also thank A. M. Nakamura, H. Mizutani, T. Tanigawa, and N. K. Mitani for fruitful discussions, and V. Shuvalov and C. Hamann as reviewers and N. Artemieva as AE for helpful comments. This work was supported in part by a Grant-in-Aid for Young Scientists B (no.22740295) from the Japan Society for the

**Table 1.** Experimental conditions and results.

| Shot No. | Sample | $h$ [a] (mm) | Laser Energy (kJ) | Laser spot diameter (mm) | $P_0$ [b] (GPa) | Distance to I-II [c] (mm) | Distance to II-III [d] (mm) | $P_s$ [f] (GPa) |
|---|---|---|---|---|---|---|---|---|
| 35245 | olivine | 0.2 | 0.69 | 0.5 | $0.9 \times 10^2$ | - [e] | 1.06 | $0.8 \times 10^2$ |
| 33772 | olivine | 0.4 | 1.3 | 0.5 | $1.5 \times 10^2$ | 0.48 | 1.38 | $1.0 \times 10^2$ |
| 33757 | olivine | 0.4 | 2.0 | 0.5 | $1.9 \times 10^2$ | 0.58 | 1.47 | $1.3 \times 10^2$ |
| 33756 | olivine | 0.6 | 4.3 | 0.5 | $3.2 \times 10^2$ | 0.91 | 2.25 | $1.1 \times 10^2$ |
| 34831 | quartz | 0.5 | 2.7 | 0.4 | $3.1 \times 10^2$ | 0.82 | 1.50 | $0.9 \times 10^2$ |
| 34829 | quartz | 0.6 | 4.4 | 0.4 | $4.3 \times 10^2$ | 1.0 | 1.62 | $1.1 \times 10^2$ |

[a] $h$: the thickness of Ti sheet.

[b] $P_0$: Laser ablation pressure.

[c] Distance from the Ti surface to the boundary between the regions I and II.

[d] Distance from the Ti surface to the boundary between the regions II and III.

[e] There is no evidence of melting.

[f] $P_s$: Peak pressure in samples estimated from numerical simulations.

# REFERENCES

Ahrens T. J. and O'Keefe J. D. 1972. Shock melting and vaporization of lunar rocks and minerals. *The Moon* 4: 214-249.

Akins J. A. and Ahrens T. J. 2002. Dynamic compression of $SiO_2$: A new interpretation. *Geophysical Research Letters* 29: 31-1-31-4.

Amsden A., Ruppel H., and Hirt C. 1980. SALE: A simplified ALE computer program for fluid flow at all speeds, Los Alamos National Laboratories Report, LA-8095:101p.

Atzeni S. and Meyer-Ter-Vehn J. 2004. The physics of inertial fusion (Beam Plasma Interaction, Hydrodynamics, Hot Dense Matter), Oxford University Press.

Collins G. S., Melosh H. J., and Ivanov B. A. (2004), Modeling damage and deformation in impact simulations. *Meteoritics and Planetary Science* 39: 217–231.

Ebert M., Hecht L., Deutsch A., and Kenkmann T. 2013. Chemical modification of projectile residues and target material in a MEMIN cratering experiment. *Meteoritics and Planetary Science* 48: 134-149.

Fabbro R., Max C., and Fabre E. 1985. Planar laser-driven ablation: Effect of inhibited electron thermal conduction. *Physics of Fluids* 28: 1463-1481.

Ferriére L., Koeberl C., Ivanov B. A., and Reimold, W. U. 2008. Shock metamorphism of Bosumtwi

impact crater rocks, shock attenuation, and uplift formation. *Science* 322: 1678-1681.

Holland K. G. and Ahrens T. J. 1997. Melting of $(Mg,Fe)_2SiO_4$ at the core-mantle boundary of the Earth. *Science* 275: 1623-1625.

Ivanov B. A., Deniem D., and Neukum G. 1997. Implementation of dynamic strength models into 2-D hydrocodes: Applications for atmospheric breakup and impact cratering, *International Journal of Impact Engineering* 20: 411–430.

Johnson G. R. and Cook W. H. 1983. A constitutive model and date for metals subjected to large strains, high strain rates and high temperatures. Seventh International Symposium on Ballistics, Hague.

Kadono T., Sakaiya T., Hironaka Y., Otani K., Sano T., Fujiwara T., Mochiyama T., Kurosawa K., Sugita S., Sekine Y., Nishikanbara W., Matsui T., Ohno S., Shiroshita A., Miyanishi K., Ozaki N., Kodama R., Nakamura A. M., Arakawa M., Fujioka S., and Shigemori K. 2010. Impact experiments with a new technique for acceleration of projectiles to a velocity higher than Earth's escape velocity 11.2 km/s. *Journal of Geophysical Research* 115: E04003, doi:10.1029/2009JE003385.

Kadono T., Shigemori K., Sakaiya T., Hironaka Y., Sano T., Watari T., Otani K., Fujiwara T., Mochiyama T., Nagatomo H., Fujioka S., Nakamura A. M., Arakawa M., Sugita S., Kurosawa K., Ohno S., and Matsui T. 2012. Flyer acceleration by high-power laser and impact experiments at velocities higher than 10 km/s. In *the Proceedings of 17th APS Topical Conference on Shock Compression of Condensed Matter*, pp. 847-850, AIP Conf. Proc. 1426 (Eds., Mark L. Elert, William T. Buttler, John P. Borg, Jennifer L. Jordan, Tracy J. Vogler).

Kieffer S. W. and Simonds C. H. 1980. The role of volatiles and lithology in the impact cratering process. *Reviews of Geophysics and Space Physics* 18: 143-181.

Kowitz A., Schmitt R. T., Reimold W. U., and Hornemann U. 2013a. The first MEMIN shock recovery experiments at low shock pressure (5-12.5 GPa) with dry, porous sandstone. *Meteoritics and Planetary Science* 48: 99-114, doi: 10.1111/maps.12030.

Kowitz A., Güldemeister N., Reimold W. U., Schmitt R. T., and Wünnemann K. 2013b. Diaplectic quartz glass and SiO2 melt experimentally generated at only 5 GPa shock pressure in porous sandstone: Laboratory observations and meso-scale numerical modeling. *Earth and Planetary Science Letters* 384: 17-26.

Kurosawa K., Sugita S., Kadono T., Shigemori K., Hironaka Y., Otani K., Sano T., Shiroshita A., Ozaki N., Miyanishi K., Sakaiya T., Sekine Y., Tachibana S., Nakamura K., Fukuzaki S., Ohno S., Kodama R., and Matsui T. 2010. In-situ spectroscopic observation of silicate vaporization due to > 10 km/s impacts using laser driven projectiles. *Geophysical Research Letters* 37: L23203, doi:

10.1029/2010GL045330.

Kurosawa K., Kadono T., Sugita S., Shigemori K., Sakaiya T., Hironaka Y., Ozaki N., Shiroshita A., Cho Y., Tachibana S., Vinci T., Ohno S., Kodama R., and Matsui T. 2012a. Shock-induced silicate vaporization: The role of electrons. *Journal of Geophysical Research* 117: E04007, doi:10.1029/2011JE004031.

Kurosawa K. Kadono T., Sugita S., Shigemori K., Hironaka Y., Sano T., Sakaiya T., Ozaki N., Shiroshita A., Ohno S., Cho Y., Hamura T., Fujioka S., Tachibana S., Vinci T., Kodama R., and Matsui T. 2012b. Time-resolved spectroscopic observations of shock-induced silicate ionization. In *the Proceedings of 17th APS Topical Conference on Shock Compression of Condensed Matter*, pp. 855-858, AIP Conf. Proc. 1426 (Eds., Mark L. Elert, William T. Buttler, John P. Borg, Jennifer L. Jordan, Tracy J. Vogler).

Melosh H. J. 1989. Impact cratering: A geologic process, Oxford University Press, New York.

Melosh H. J., Ryan E. V., and Asphaug E. 1992. Dynamic fragmentation in impacts: Hydrocode simulation of laboratory impacts. *Journal of Geophysical Research* 97: 14,735–14,795.

Mitani N. K. 2003. Numerical simulations of shock attenuation in solids and reevaluation of scaling law. *Journal of Geophysical Research* 108: 5003, doi:10.1029/2000JE001472.

<p>1+1</p>

Mizutani H., Takagi Y., and Kawakami S. 1990. New scaling laws on impact fragmentation. *Icarus* 87: 307-326.

Ohno S. Kadono T., Kurosawa K., Hamura T., Sakaiya T., Shigemori K., Hironaka Y., Sano T., Watari T., Otani K., Matsui T., and Sugita S. 2014. Production of sulphate-rich vapour during the Chicxulub impact and implications for ocean acidification. *Nature Geoscience* 7: 279-282.

Ohtani E., Kimura Y., Kimura M., Takata T., Kondo K., and Kubo T. 2004. Formation of high-pressure minerals in shocked L6 chondrite Yamato 791384: constraints on shock conditions and parent body size. *Earth and Planetary Science Letters* 227: 505-515.

Pierazzo E., Vickery A. M., and Melosh H. J. 1997. A Reevaluation of impact melt production. *Icarus* 127: 408-423.

Reimold W. U. and Stöffler D. 1978. Experimental shock metamorphism of dunite. *Proceedings of Lunar Planetary Science Conference* 9th: 2805-2824.

Stöffler D. 1972. Deformation and transformation of rock-forming minerals by natural and experimental shock processes I. Behavior of minerals under shock compression. *Fortschritte der Mineralogie* 49: 50-113.

Stöffler D. and Langenhorst F. 1994. Shock metamorphism of quartz in nature and experiment: I. Basic

observation and theory. *Meteoritics* 29: 155-181.

Stöffler D., Gault D. E., Wedekind J., and Polkowski G. 1975. Experimental hypervelocity impact into quartz sand: Distribution and shock metamorphism of ejecta. *Journal of Geophysical Research* 80: 4062-4077.

Sugita S., Kurosawa K., and Kadono T. 2012. A semi-analytical on-Hugoniot EOS of condensed matter using a linear Up − Us relation. In *the Proceedings of 17th APS Topical Conference on Shock Compression of Condensed Matter*, pp. 895-898, AIP Conf. Proc. 1426 (Eds., Mark L. Elert, William T. Buttler, John P. Borg, Jennifer L. Jordan, Tracy J. Vogler).

Syono Y., Goto T., Takei H., Tokonami M., and Nobugai K. 1981a. Dissociation reaction in Forsterite under shock compression. *Science* 214: 177-179.

Syono Y., Goto T., Sato J., and Takei H. 1981b. Shock compression measurements of single-crystal forsterite in the pressure range 15-93 GPa. *Journal of Geophysical Research* 86: 6181-6186.

Takasawa S., Nakamura A. M., Kadono T., Arakawa M., Dohi K., Ohno S., Seto Y., Maeda M., Shigemori K., Hironaka Y., Sakaiya T., Fujioka S., Sano T., Otani K., Watari T., Sangen K., Setoh M., Machii N., and Takeuchi T. 2011. Silicate dust size distribution from hypervelocity collisions: Implications for dust production in debris disks. *Astrophysical Journal Letters* 733: L39-L42.

Thompson S. L. and Lauson H. S. 1972. Improvements in the Chart-D radiation hydrodynamic code III: Revised analytical equation of state. pp. SC-RR-71 0714, 119 pp., Sandia Laboratories, Albuquerque, NM.

Tillotson J. H. 1962. Metallic equations of state for hypervelocity impact. Technical Report GA-3216, General Atomic Report.

Tschauner O., Asimow P. D., Kostandova N., Ahrens T. J., Ma C., Sinogeikin S., Liu Z., Fakra S., and Tamura N. 2009. Ultrafast growth of wadsleyite in shock-produced melts and its implications for early solar system impact processes. *Proceedings of the National Academy of Sciences* 106: 13691-13695.

Wünnemann K., Collins G. S., and Melosh H. J. 2006. A strain-based porosity model for use in hydrocode simulations of impacts and implications for transient crater growth in porous targets. *Icarus* 180: 514–527.

**Figures and Captions**

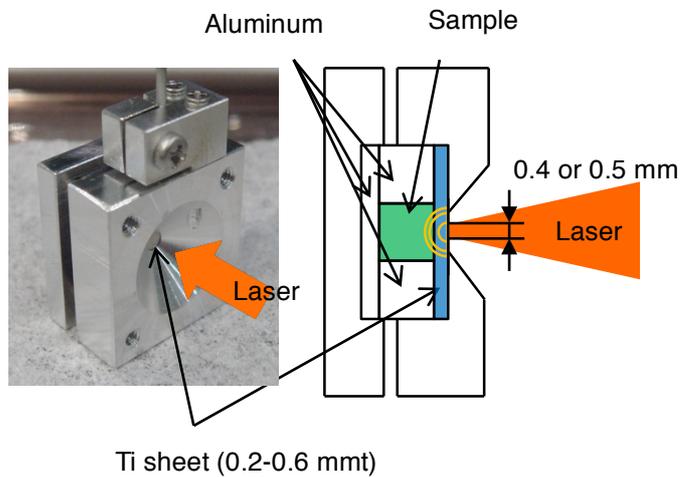

**Fig. 1.** Target setup for the recovery experiments. Cubic olivine or quartz samples (sides 3 mm) were surrounded by Al cases, except for the front surface, which was covered with a Ti sheet (thickness 0.2-0.6 mm) to prevent the ejection of shocked samples from the Al cases. The Ti sheet was irradiated with a laser of spot diameter of 0.4 or 0.5 mm, generating a shock wave in Ti that propagated into the sample. The samples were recovered after the shot.

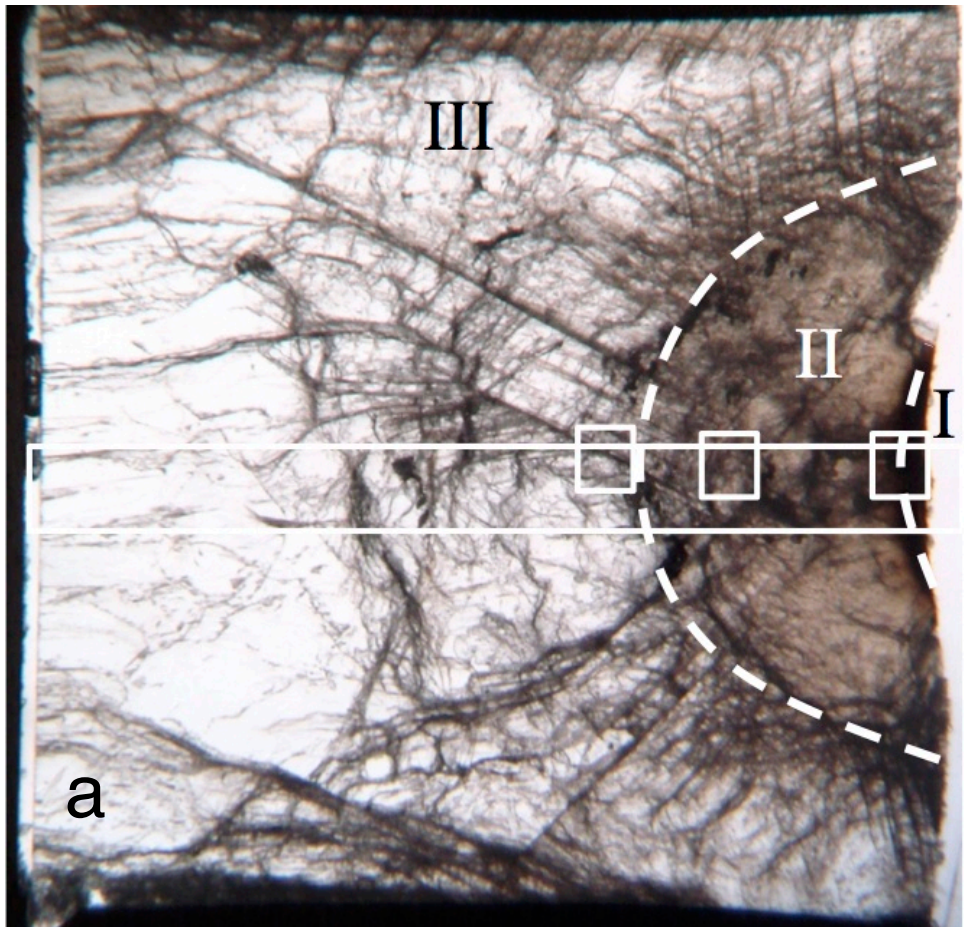
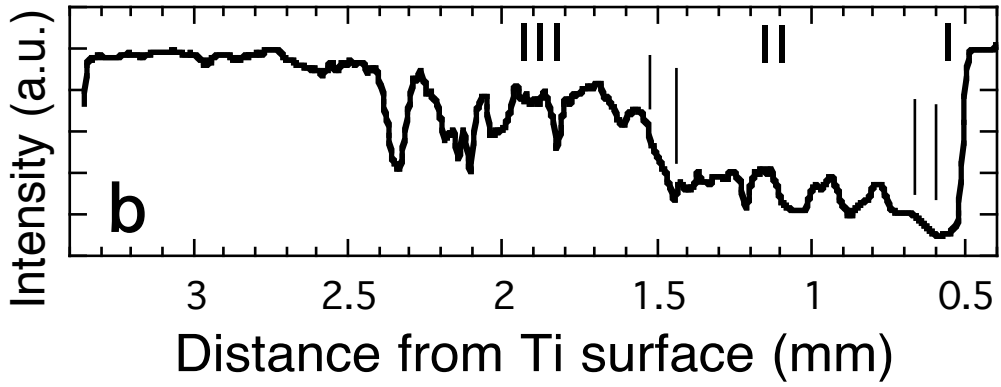
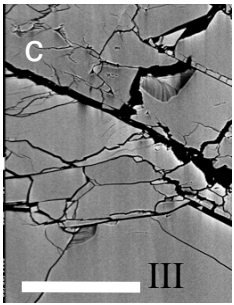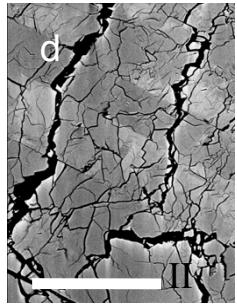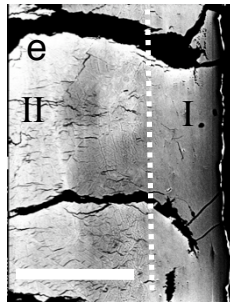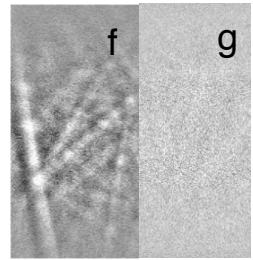

**Fig. 2.** (a) An optical (transmitted light) microscope image of a slice of the recovered sample of shot no. 33757. The shock wave arrived from the right. The right-hand edge of the sample was slightly lost about 85 μm from the original surface. Broken lines divide (I) the melt, (II) plastic, and (III) elastic region. (b) Intensity distribution in the large white rectangle in (a) as a function of the distance from the Ti surface (vertically integrated). The boundaries between I and II, and II and III, are defined as the middle points between adjacent thin-vertical lines. (c) SEM image of the area delineated by the rectangle in (a) farthest to the surface in Region III. Fewer cracks (fragments) appear; cracks tend to align. (d) SEM image of Region II. Many cracks oriented in various directions are visible. (e) SEM image of the area delineated by the rectangle closest to the surface including the boundary between Regions I and II (white dotted line). Regions I and II are distinguished by few and many cracks, respectively. (f) and (g) EBSD images obtained from the left-hand side (corresponding to Region II) and the right-hand side (Region I) of the image (e), respectively. Regions I and II are distinguished by a clearer pattern in the former. The white horizontal bars in (c), (d), and (e) denote a scale of 100 μm.

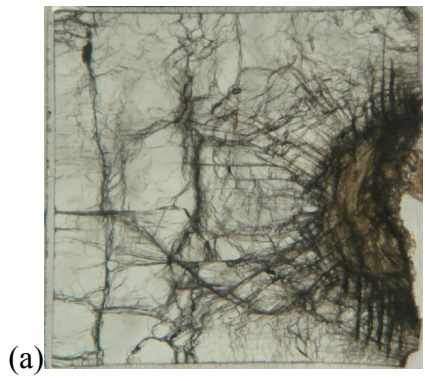 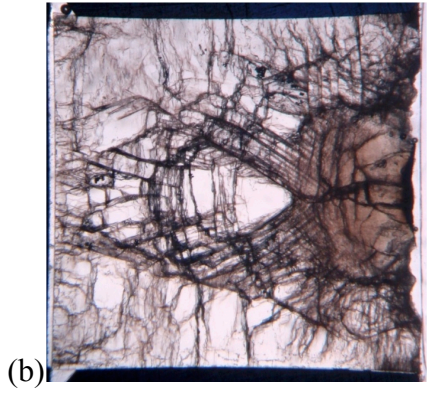

(a) (b)

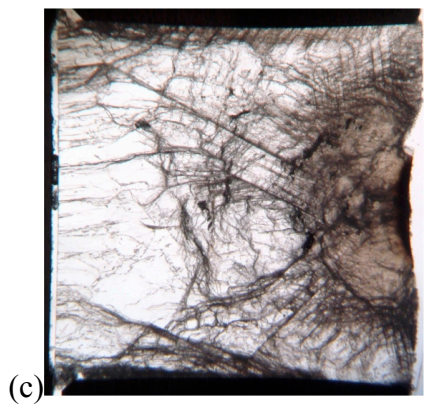 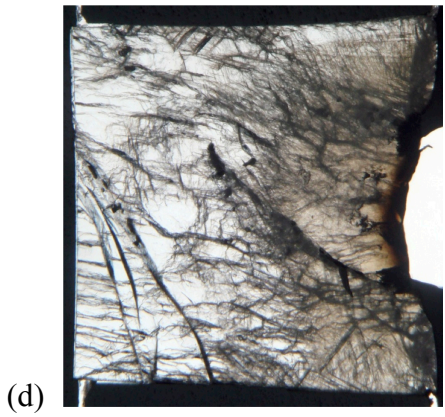

(c) (d)

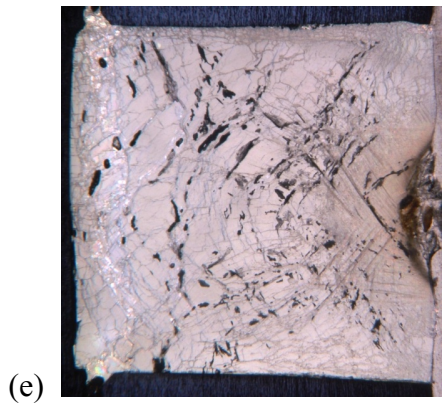 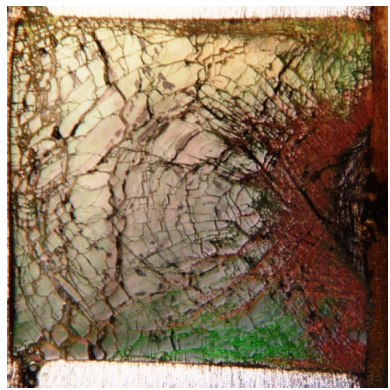

(e)

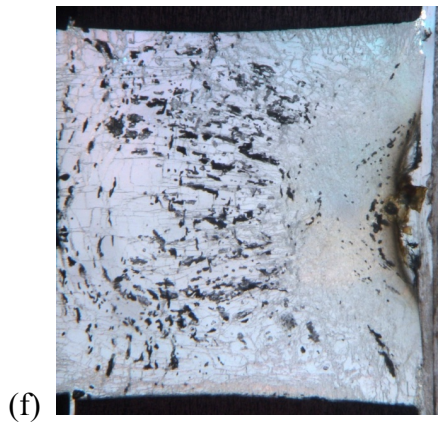 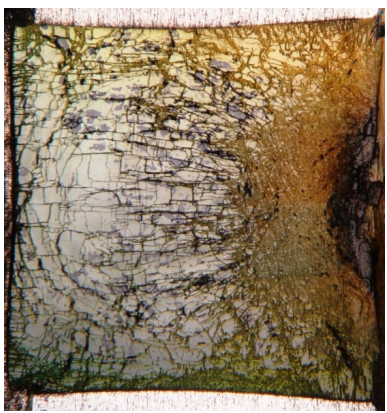

(f)

**Fig. 3.** Optical (transmitted light) microscope images of slices of the recovered samples: (a) olivine (#35245), there seems no Region I, (b) olivine (#33772), (c) olivine (#33757; the same as Fig. 2(a)), (d) olivine (#33756), (e) quartz (#34831), and (f) quartz (#34829). For the sections of the quartz samples in (e) and (f), not only the transmitted−light image (the left-hand side) but also the reflected-light image (the right-hand side) are shown. The length of each image is 3 mm. Laser irradiated from the right-hand side.

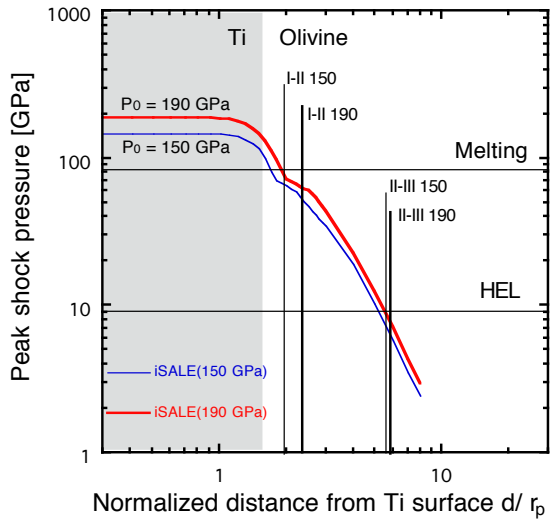

**Fig. 4.** Shock pressure as a function of normalized distance ($d/r_p$). Numerical results by iSALE with strength are shown for the initial pressure $P_0$ of 150 and 190 GPa as a thin curve and a bold one, respectively. The horizontal lines indicate the melting pressure and HEL of olivine. The positions of the boundaries between I and II and II and III obtained from the recovered samples are indicated with the vertical lines labeled I-II and II-III for $P_0$ of 150 GPa (thin lines) and 190 GPa (bold ones). The number following these labels means $P_0$.

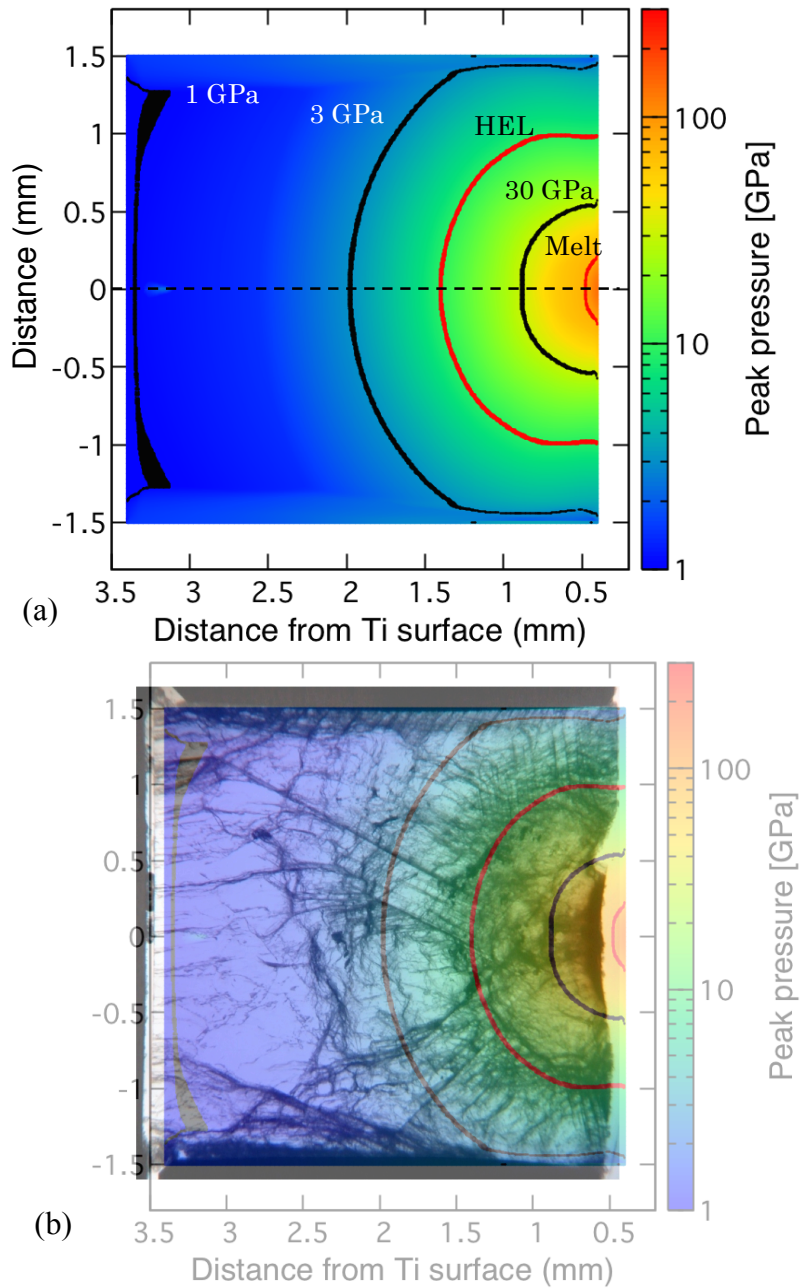

**Fig. 5.** (a) Pressure contour in the olivine target for $P_0$ of 190 GPa obtained by iSALE-2D calculation in axis-symmetry. The target is surrounded by Al cases except for one surface where a Ti sheet covers and a projectile impacts (the right-hand side of the figure). Ti and Al are not shown in this figure. The horizontal axis is the distance

from the Ti surface in mm. The color indicates the pressure experienced as shown in the right-hand side in log-scale. Two red lines means the pressures of the onset of melting and HEL. The horizontal broken line indicates the central axis, along which the projectile impacts from the right. (b) The same pressure contour by iSALE-2D as (a) superimposed on the image of the recovered olivine sample in Fig. 2a.

**Electronic Appendix**

**Table A1.** General setup parameters for the iSALE calculations

| | |
|---|---|
| Computational geometry | Cylindrical coordinate |
| Number of computational cells in R direction | 720 |
| Number of computational cells in Z direction | 820 |
| Number of cells for samples in R direction | 300 |
| Number of cells for samples in Z direction | 300 |
| Number of cells for the Al case in R direction | 400[a] |
| Grid spacing (μm/grid) | 5 |
| Artificial viscosity[b] $a_1$ | 0.24 |
| Artificial viscosity $a_2$ | 1.2 |

a. This value was chosen to stand off the wave interaction with the rarefaction wave from the surface of the outside of the aluminum case. The actual thickness of the aluminum case used in the experiments was much thicker than this value. Thus, the effect of the rarefaction wave from the surface of the outside of the aluminum case can be also neglected under the experimental conditions.

b. The artificial viscosity is introduced into the iSALE to capture a shock in computation and to dampen a numerical oscillation behind the shock. This leads to shock smearing with the full width at half maximum of ~3 cells [e.g., *Johnson et al.*, 2014]. The grid spacing of 5 μm/grid was chosen to minimize the effect of the shock smearing on the calculated spatial distribution of the peak pressure. The grid spacing used in this study, 5 μm/grid, provides the number of cells for titanium flyer in Z direction with >24 cells.

**Table A2.** The list of input parameters for minerals in iSALE calculations.

| | Dunite | Quartz |
|---|---|---|
| EOS type | ANEOS[a] | ANEOS[b] |
| Strength model | Rock[c] | Rock[c] |
| Poisson ratio | 0.25[d] | 0.25[d] |
| Melting temperature (K) | 1373[d] | 1750[e] |
| Thermal softening coefficient | 1.1[d] | 1.1[f] |
| Simon parameter A (GPa) | 1.52[d] | 1.8[e] |
| Simon parameter C | 4.05[d] | 3.4[e] |
| Cohesion (Undamaged) (MPa) | 5.07[d] | 1000[f] |
| Cohesion (Damaged) (kPa) | 10[d] | 10[f] |
| Internal friction (Undamaged) | 1.58[d] | 1.2[f] |
| Internal friction (Damaged) | 0.63[d] | 0.6[f] |
| Limiting strength (GPa) | 3.26[d] | 3.5[f] |
| Minimum failure strain | $10^{-4}$ [f] | $10^{-4}$ [f] |
| A constant for the damage model | $10^{-11}$ [f] | $10^{-11}$ [f] |
| A threshold pressure for the damage model (MPa) | 300[f] | 300[f] |

a. *Benz et al.* (1989)

b. *Melosh* (2007)

c. The detailed description of the strength model for rocks can be found in *Collins et al.* (2004).

d. *Johnson et al.* (2015)

e. *Wünnemann et al.* (2008)

f. Typical values for minerals are employed.

**Table A3.** The list of input parameters for metals in iSALE calculations.

| | Titanium | Aluminum |
|---|---|---|
| EOS type | Tillotson[a] | Tillotson[a] |
| Strength model | Johnson-Cook[b] | Johnson-Cook[b] |
| Poisson ratio | 0.33[c] | 0.33[c] |
| Melting temperature at 1 atm | 1933[d] | 1063[e] |
| Isochoric specific heat (J/K/kg) | 521[f] | 896[f] |
| Simon parameter A (GPa) | 1.5[d] | 8.8[e] |
| Simon parameter C | 5.3[d] | 2.3[e] |
| Johnson-Cook parameter A (MPa) | 880[g] | 321[b] |
| Johnson-Cook parameter B (MPa) | 695[g] | 114[b] |
| Johnson-Cook parameter N | 0.36[g] | 0.42[b] |
| Johnson-Cook parameter C | 0.04[g] | 0.002[b] |
| Johnson-Cook parameter M | 0.8[g] | 1.34[b] |
| Reference temperature (K) | 293 | 293 |

a. *Tillotson* (1962)
b. The detailed description of the model can be found in *Johnson and Cook* (1983)
c. A typical value for metals was employed.
d. We fitted the melting curve by *Kerley* (2003) with the Simon equation up to 100 GPa [e.g., *Wünnemann et al.*, 2008]. The melting curve calculation using the Simon equation is supported in the iSALE. Note that the melting curve above 100 GPa by *Kerley* (2003) deviates from the calculated value of the Simon equation with the parameters shown here.
e. *Zhang et al.* (2014)
f. The isochoric specific heat $C_v$ was assumed to be the Dulong-Petit limit.
g. *Dorogoy and Rittel* (2009). Note that we used the parameters for a titanium alloy $Ti_6Al_4V$ instead of pure titanium because the parameters of Johnson-Cook strength model for pure Ti have not been established. To access the validity this treatment, we conducted one more calculation without strength in the titanium plate on the surface of the targets and the projectile. Figure A4 shows the effect of strength of Ti on the peak pressure distribution in dunite sample. The initial shock pressure $P_o$ was set to 190 GPa in this calculation, which is the same as shown in Figure 4 in the main text. This comparison clearly shows that the strength of Ti does not largely affect the pressure distribution in the samples. The shocked Ti around the central axis is mostly melted, resulting in drastic decrease in the strength due to the effect of thermal softening.

Figures and captions

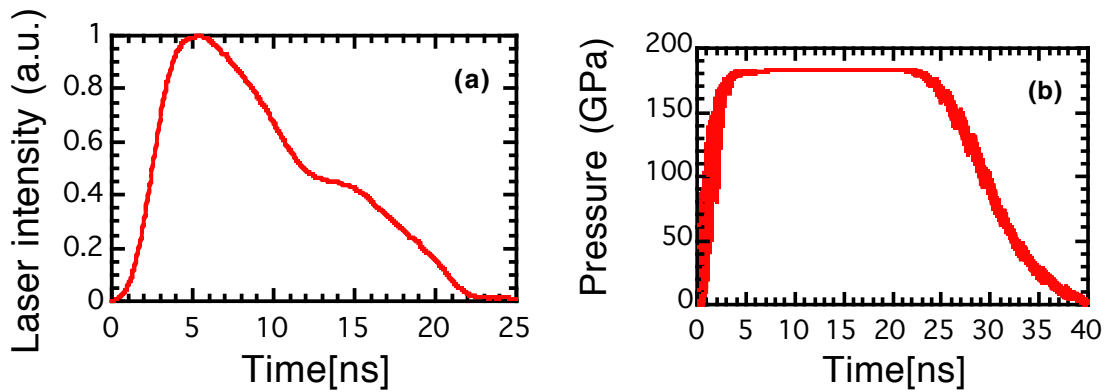

**Figure A1.** (a) Typical laser intensity profile in the experiments and (b) initial pressure one caused by virtual projectile impact in the iSALE numerical simulation. The duration of the flat area depends on the dimension of the virtual projectile and affects the positions of melt and HEL. The dependence of the positions on the projectile thickness is shown in Fig. A5.

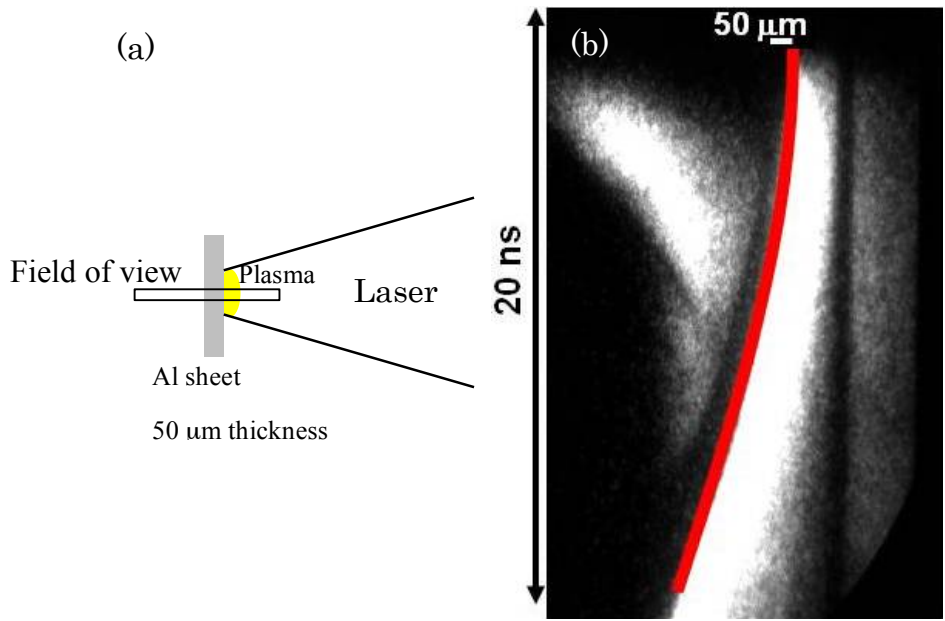

**Figure A2.** Calibration experiment. (a) Experiment setup. A thin (50 μm in thickness) aluminum (Al) sheet was irradiated by the laser at the same pulse width and wavelength as those in our experiments. Laser energy and spot diameter were 2.3 kJ and 0.5 mm in diameter, respectively. (b) A streak camera image of the calibration experiment. The vertical axis is time, proceeding from the top to the bottom. The horizontal axis is space; the horizontal white bar at the top of the figure is the original position of the Al sheet and the length indicates 50 μm. The laser irradiated from the right and the bright parts just on the right side of the Al sheet is the ablation plasma. At each time, the edge of the brightest part of plasma corresponds to the surface of the Al sheet indicated by the red curve (the Al sheet moves to the left). The curve is approximately regarded to be parabolic; the Al sheet was accelerated at an approximately constant acceleration, that is, an approximately constant pressure, during the laser irradiation. We define a constant average (effective) ablation pressure $P_0$ as $ma/S = \rho d a$, where $S$, $m$, $\rho$, $d$, and $a$ are laser irradiation spot area, the mass, density, and thickness of the Al sheet flyer, and acceleration, respectively.

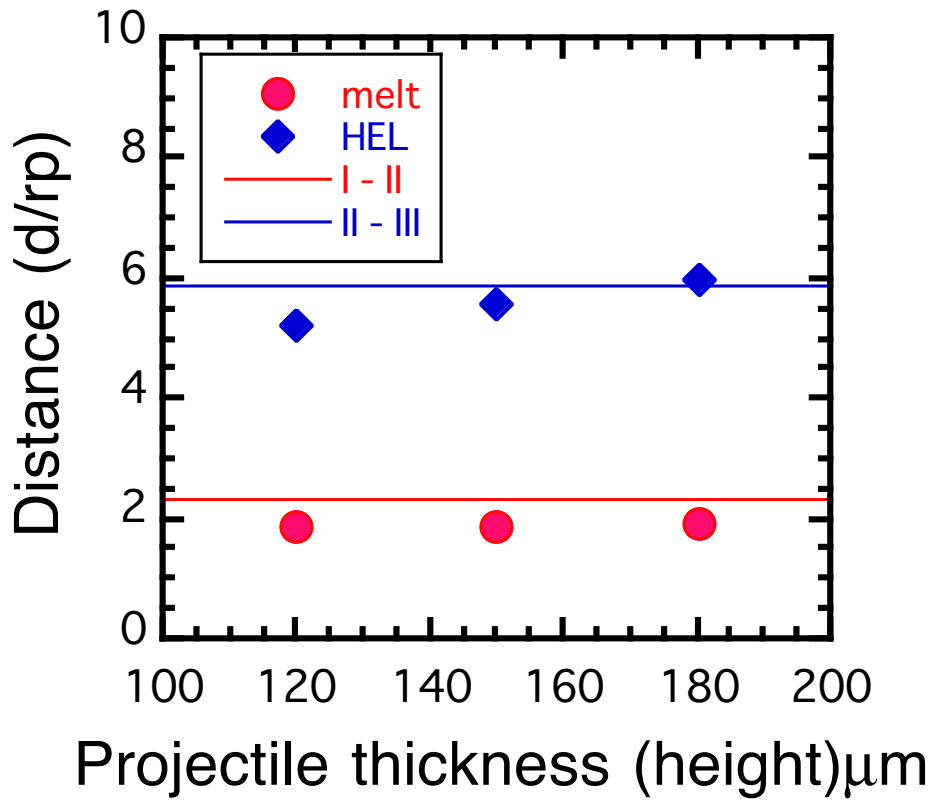

**Figure A3.** The positions of melt and HEL (= the distance from the Ti surface normalized by the projectile radius $r_p$) obtained by iSALE as a function of projectile thickness (height). The corresponding experiment is the case of $P_0 = 190$ GPa. Horizontal lines indicate the boundaries between I and II (red) and II and III (blue) obtained in the recovered sample. The positions of melt and HEL in the simulations increase with the projectile thickness, but the slope is gradual. The result shown in Fig. 4 for $P_0 = 190$ GPa is obtained with a projectile thickness of 150 μm.

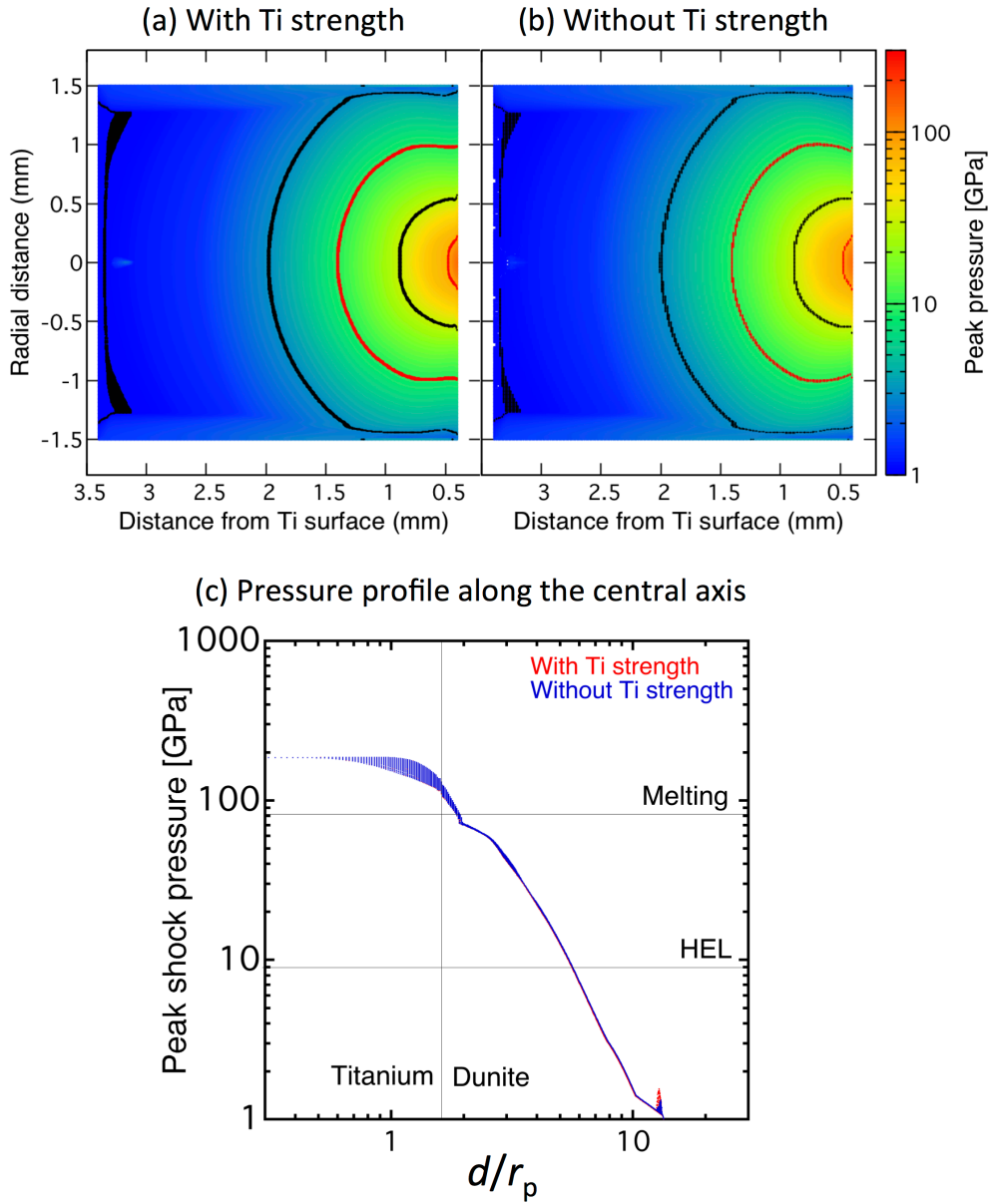

**Figure A4**. The effect of the strength of the titanium. (a) The pressure contour with Johnson-Cook strength model in which the parameters for the titanium alloy were used as Ti. This figure is the same as Figure 4 in the main text. (b) The pressure contour without strength model for Ti. (c) The peak shock pressure along the central axis as a function of $d/r_p$. Two lines with and without strength for Ti mostly overlap.

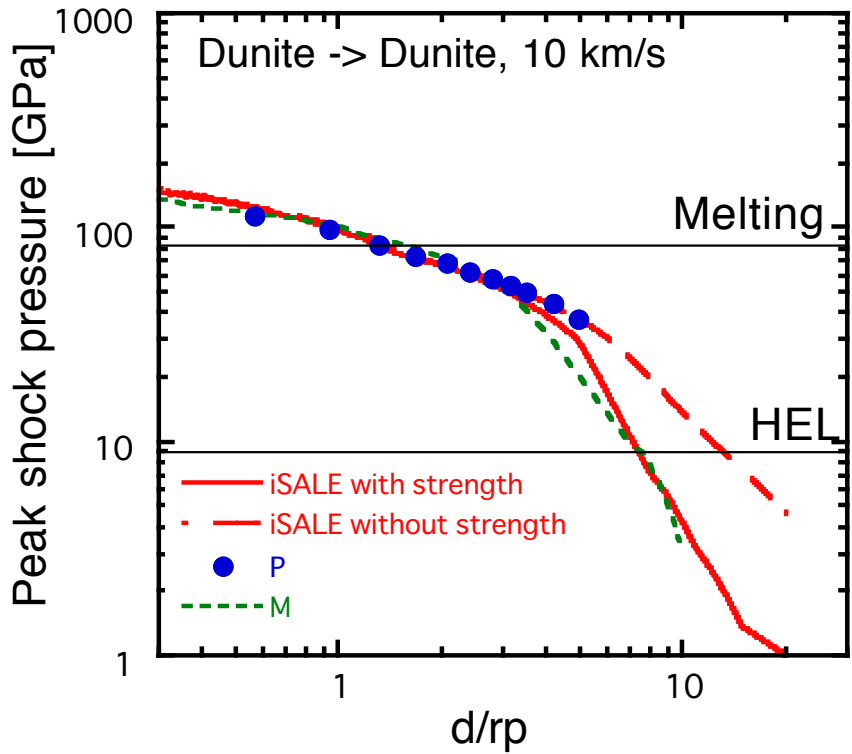

**Figure A5.** A pressure profile along the central axis as a function of the distance normalized by projectile radius ($d/r_p$). We plot the results by iSALE for an impact velocity of 10 km/s with and without strength as a bold curve and a broken one, respectively. The number of cells for projectile radius (CPPR) was set to 50. For comparison, we show two previous numerical results for dunite - dunite impact at 10 km/s obtained by Pierazzo et al. (1997) (denoted by "P") and Mitani (2003) (denoted by "M"). Since the code by Mitani (2003) includes a constituent equation specified for a pressure around HEL, the pressure decreases more rapidly than the result by Pierazzo et al. (1997) at several tens GPa. The result by iSALE without strength is almost the same profile as the result (P). The result by iSALE with strength shows that the pressure decreases more quickly and consistent with that by Mitani (2003). The horizontal lines indicate the melting pressure and HEL of olivine.

(a) #35245 Olivine ($P_0 = 0.9 \times 10^2$ GPa)

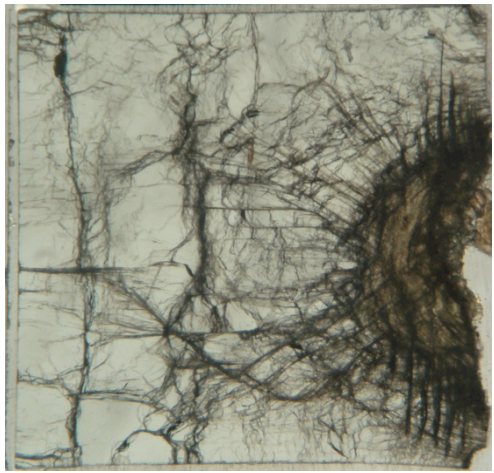

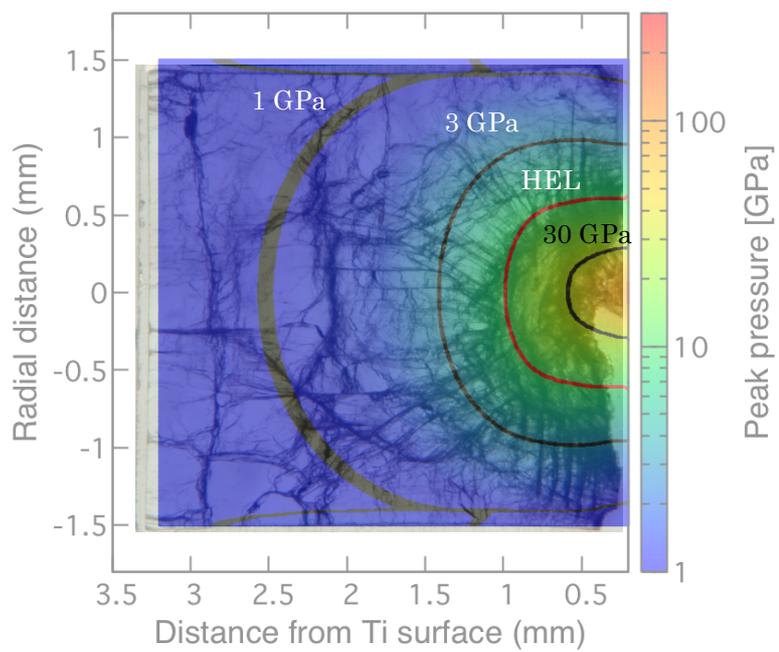

(b) #33772 Olivine ($P_0 = 1.5 \times 10^2$ GPa)

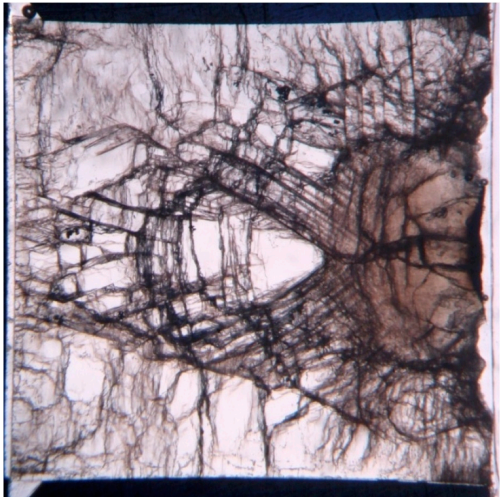

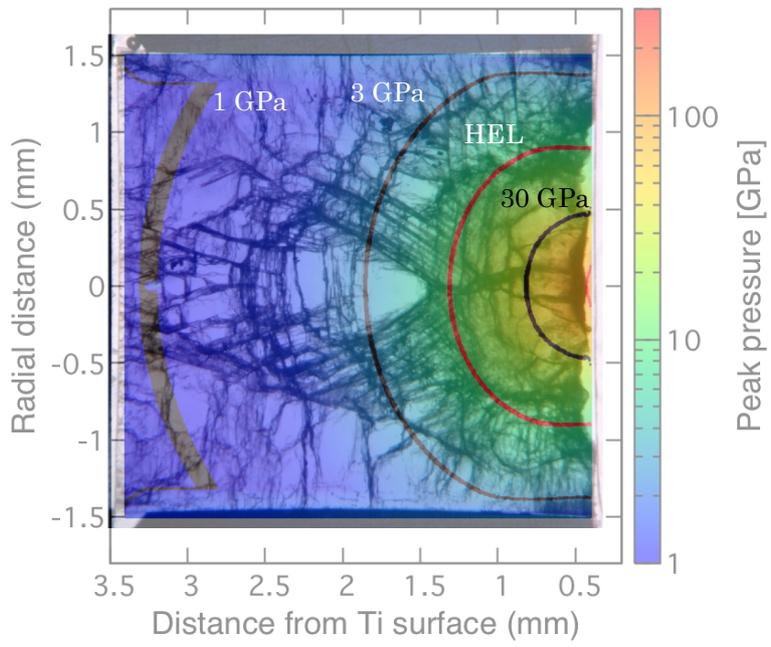

(c) #33757 Olivine ($P_0 = 1.9 \times 10^2$ GPa)

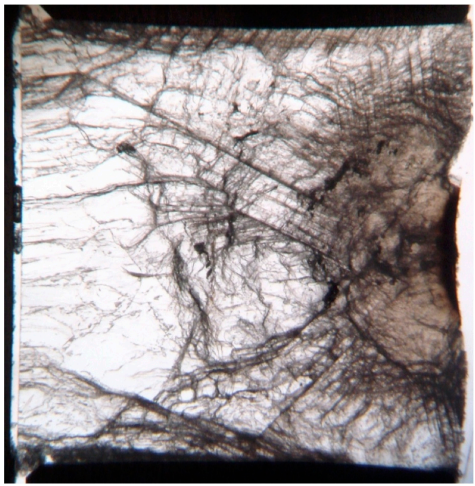

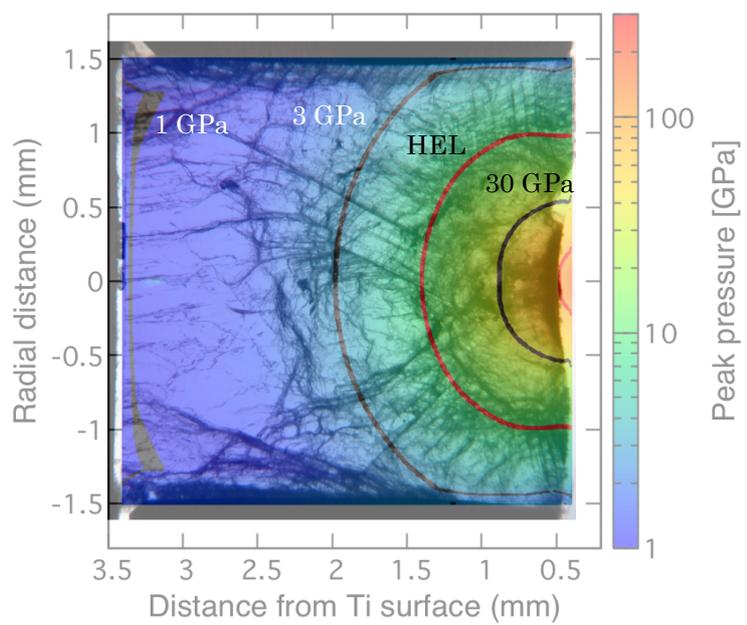

(d) #33756 Olivine ($P_0 = 3.2 \times 10^2$ GPa)

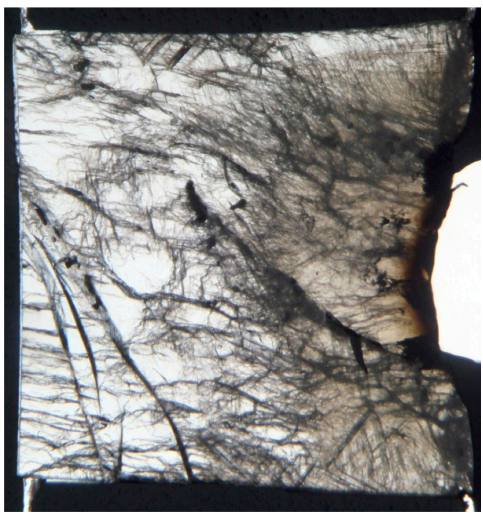

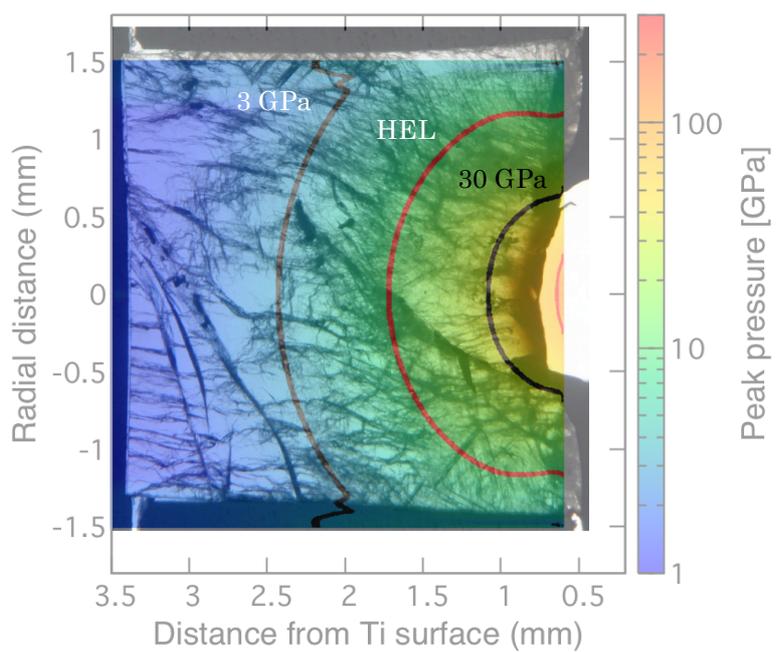

(e) #34831 Quartz ($P_0 = 3.1 \times 10^2$ GPa)

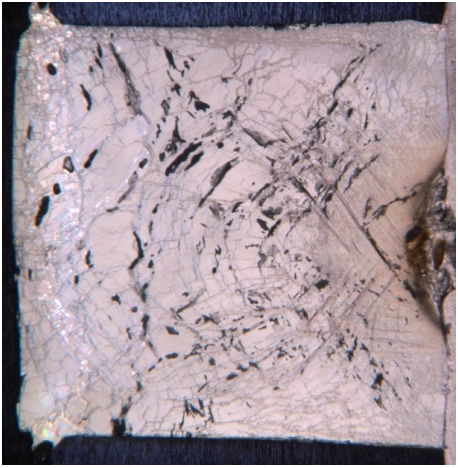
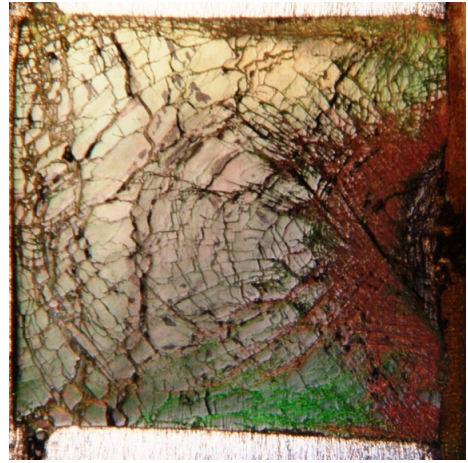
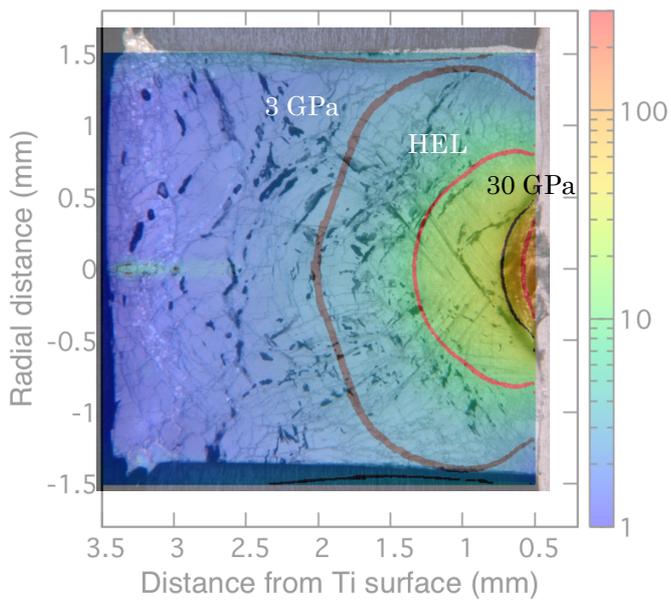
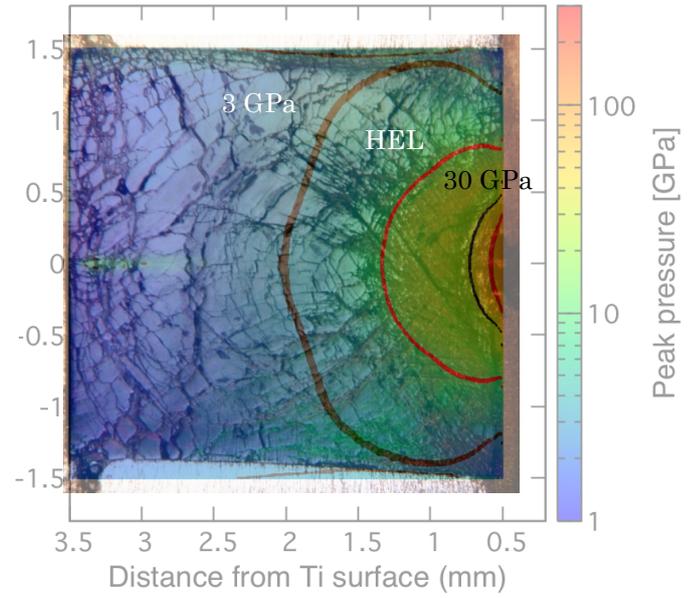

(f) #34829 Quartz ($P_0 = 4.3 \times 10^2$ GPa)

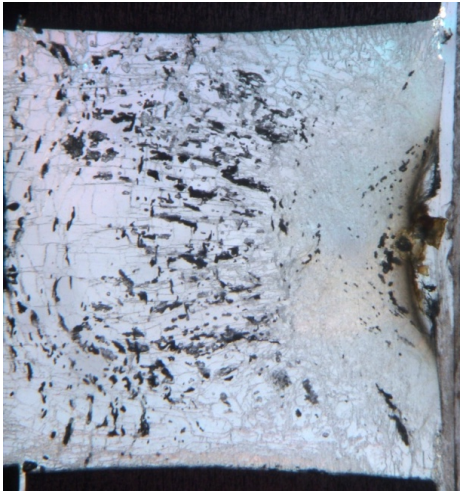
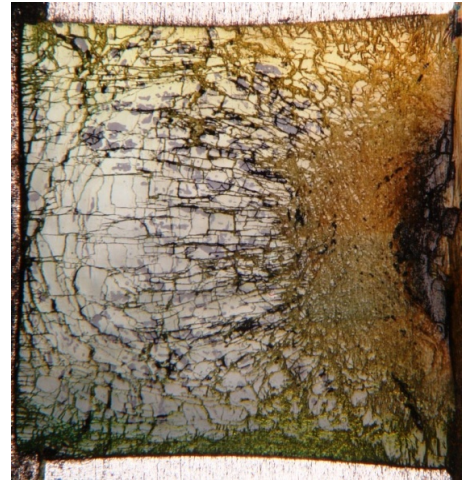
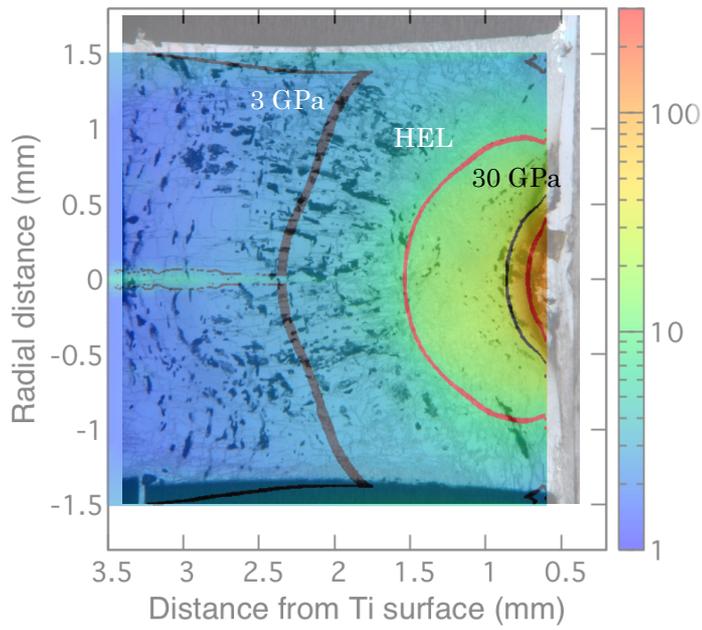
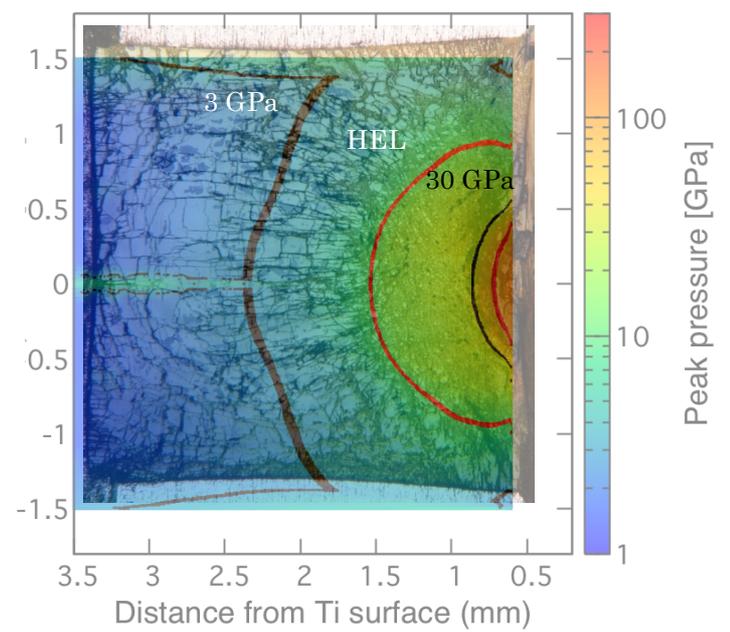

**Figure A6.** Optical (transmitted light) microscope images of thin-sections of the recovered samples: (a) olivine (#35245), there are no melt region, (b) olivine (#33772), (c) olivine (#33757; the same as Fig. 2(a) and Fig. 4), (d) olivine (#33756), (e) quartz (#34831), and (f) quartz (#34829). For the sections of the quartz samples in (e) and (f), not only the transmitted−light image (the left-hand side) but also the reflected-light image (the right-hand side) are shown. The length of each image is 3 mm. Laser irradiated from the right-hand side.

**References in Appendix**

Benz, W., Cameron, A. and Melosh, H. J. 1989. The origin of the Moon and the single-impact hypothesis III. *Icarus*, 81: 113-131.

Collins, G. S., Melosh, H. J., and Ivanov, B. A. 2004. Modeling and deformation in impact simulations. *Meteoritics and Planetary Science*, 39: 217-231.

Dorogoy, A. and Rittel, D. 2009. Determination of the Johnson-Cook material parameters using the SCS specimen. *Experimental Mechanics*, 49: 881-885.

Johnson, G. R. and Cook, W. H. 1983. A constitutive model and data for metals subjected to large strains, high strain rates and high temperatures, *Seventh International Symposium on Ballistics, Hague*.

Johnson, B. C., Bowling, T. J., and Melosh, H. J. 2014. Jetting during vertical impacts of spherical projectiles. *Icarus*, 238, 13-22.

Johnson, B. C., Minton, D. A., Melosh, H. J. and Zuber, M. T. 2015. Impact jetting as the origin of chondrules. *Nature*, 517: 339-341.

Kerley, G. I. 2003. Equations of state for Titanium and Ti6A14V Alloy. *Sandia report*, SAND 2003-3785.

Melosh, H. J. 2007. A hydrocode equation of state for $SiO_2$. *Meteoritics and Planetary Science,* 42: 2079-2098.

Mitani, N. K., 2003. Numerical simulations of shock attenuation in solids and reevaluation of scaling law. *Journal of Geophysical Research* 108: 5003, doi:10.1029/2000JE001472.

Pierazzo, E., Vickery, A. M., Melosh, H. J., 1997. A Reevaluation of impact melt production. *Icarus* 127: 408-423.

Tillotson, J. H. 1962. Metallic equations of state for hypervelocity impact. *Technical Report GA-3216*, General Atomic Report.

Wünnemann, K., Collins, G. S., and Osinski, G. R. 2008. Numerical modeling of impact melt production in porous rocks. *Earth and Planetary Science Letters*, 269: 530-539.

Zhang, W., Peng, Y. and Liu, Z. 2014. Molecular dynamics simulations of the melting curve of NiAl alloy under pressure. *AIP Advances*, 4: 057110.